\documentclass[12pt,twoside]{article}
\input{epsf.sty}

\def\mydate{26 July 1999}
\def\ignore#1{{}}

\newcommand{\beeq}{\begin{equation}}
\newcommand{\eneq}{\end{equation}}
\newcommand{\beqn}{\begin{eqnarray}}
\newcommand{\eeqn}{\end{eqnarray}}

\def\la{\raise.16ex\hbox{$\langle$} \, }
\def\ra{\, \raise.16ex\hbox{$\rangle$} }
\def\ran{\raise.16ex\hbox{$\rangle$} }

\def\next{{~~~,~~~}}

\def\psibar{ \psi \kern-.65em\raise.6em\hbox{$-$}\lower.6em\hbox{} }
\def\Psibar{ \Psi \kern-.77em\raise.6em\hbox{$-$} }
\def\Phibar{ \Phi \kern-.77em\raise.6em\hbox{$-$} }

\begin{document}

\baselineskip=12pt

{\small \hfill  UMN-TH-1803/99\\}
{\small \noindent \mydate \hfill UMN-TH-1810/99}

\baselineskip=40pt plus 1pt minus 1pt

\vskip 3cm

\begin{center}

{\Large\bf {Monopole and dyon solutions in the  Einstein Yang-Mills
theory in asymptotically anti-de Sitter space\footnote[1]{Proceedings from the talk given
at ``The eighth Canadian Conference on General Relativity and
Astrophysics'', McGill University, Montreal, Quebec, Canada, June 10-12
1999.}}}\\

\vspace{1.5cm}
\baselineskip=20pt plus 1pt minus 1pt

{\large   Yutaka Hosotani and Jefferson Bjoraker}\\
\vspace{.1cm}
{\it School of Physics and Astronomy, University of Minnesota}\\  
{\it  Minneapolis, MN 55455, U.S.A.}\\ 
\end{center}

\vskip 3.cm
\baselineskip=20pt plus 1pt minus 1pt

\begin{abstract}
Regular monopole and dyon solutions to the $SU(2)$ Einstein Yang-Mills
equations in asymptotically anti-de Sitter space are discussed. A
class of monopole solutions are shown to be stable against spherically
symmetric linear perturbations.
\end{abstract}

\section{Introduction}

Static solutions to the Einstein Yang-Mills (EYM) equations differ considerably
depending on the value of the cosmological constant. The solutions can
be separated into two families; $\Lambda \ge 0$, and $\Lambda <
0$. The solutions where $\Lambda = 0$ were discovered by Bartnik
and McKinnon (BK) \cite{BARTNIK} and their asymptotically de Sitter (dS)  analogs
($\Lambda > 0$) were discovered independently by Volkov et. al. and
Torii et. al. \cite{VOLKOV}. The BK solutions and the cosmological
extensions to them all share similar behavior and have been studied in
detail (see Ref. \cite{VOLKOV2} for a review).  
Recently,  asymptotically anti-de
Sitter ($\Lambda < 0$) black hole solutions \cite{WINSTANLEY} and soliton solutions
\cite{BJORAKER} were found which are strikingly different than the
BK type solutions. In particular, there exist asymptotically anti-de Sitter (AdS)
solitons with no nodes in the field strength which are
stable. Furthermore, in asymptotically AdS space dyon solutions are allowed.

\section{General formalism}
 
Given the spherically symmetric metric in the Schwarzschild gauge
\begin{equation}
ds^2 = -\frac{H dt^2}{p^2} +\frac{dr^2}{H} 
+r^2 (d\theta^2 + \sin^2\theta d\phi^2),
\label{GMUNU}
\end{equation}
where $H$, $p$ are functions of $t$ and $r$,
the coupled static EYM equations of motion are 
\beqn
\left(\frac{H}{p}w^{\prime}\right)^{\prime} 
&=& -\frac{p}{H}u^2w-\frac{w}{p}\frac{(1-w^2)}{r^2} 
\label{YM1}  \\
\left(r^2pu^{\prime}\right)^{\prime}& =& \frac{2p}{H}w^2u  
\label{YM2}\\
p^{\prime} &=&
-\frac{2v}{r}p\left[(w^{\prime})^2+\frac{u^2w^2p^2}{H^2}\right] 
\label{Ein1}  \\
m' &=& v\left[
\frac{(w^2-1)^2}{2r^2}+\frac{1}{2}r^2p^2(u^{\prime})^2
  +H(w^{\prime})^2+\frac{u^2w^2p^2}{H} \right] ~,
\label{Ein2}
\eeqn
where $w(r)$ and $u(r)$ are the magnetic and electric components of
the $SU(2)$ Yang-Mills fields \cite{WITTEN},
$H(r) = 1 - {2m(r)}/{r} - \Lambda r^2/3$, $v ={G}/{4\pi e^2}$
and $\Lambda$ is the cosmological constant.  We require that solutions
to Eq.'s (\ref{YM1}) to (\ref{Ein2})
are regular everywhere and have finite ADM mass $M = m(\infty)$.
The electric and magnetic charges of solutions are given by
\beeq
\pmatrix{Q_E\cr Q_M\cr}
 = {e\over 4\pi} \int dS_k \, \sqrt{-g} \, 
\pmatrix{ F^{k0}\cr \tilde F^{k0} \cr}
= \pmatrix{u_1 p_0\cr 1 - w_0^2\cr} {\tau_3\over 2}
\label{charge1}
\eneq
where $w=w_0 + w_1/r + {\rm O}(r^{-2})$ etc.

For solutions in asymptotically flat or dS space, $w(r)$ has at least one
node
\cite{VOLKOV}. The situation is quite different in  asymptotically AdS
space ($\Lambda<0$).  $H(r)$ is positive everywhere and there are
solutions where $w$ has no node.

\section{Monopole solutions}

The solutions to Eq.'s (\ref{YM1}) to (\ref{Ein2}), for $\Lambda < 0$, were
evaluated numerically \cite{BJORAKER} using the shooting method. In
the shooting method one solves Eqs.\ (\ref{YM1}) to (\ref{Ein2}) at $r=0$
in terms of two parameters, $a$ and $b$, and
`shoots' for solutions with the desired asymptotic behavior. 
$a$ and $b$ are adjustable parameters which together specify the
boundary conditions at the origin for $w$, $u$, $H$, and $p$:
$u(r)=ar+\cdots$ and $w(r) = 1-br^2+\cdots$ near $r=0$.

Purely magnetic solutions (monopoles) are found by setting $a=0$,
corresponding to $u(r)=0$. A continuum of
monopole solutions were obtained by varying the parameter $b$. The
solutions are similar to the black hole
solutions found by Ref. \cite{WINSTANLEY}, but are also regular for
all $r$.  The number of times $w$ crosses the axis depends on
the value of the adjustable shooting parameter $b$.

\begin{figure}[th]\centering
 \leavevmode 
\mbox{
\epsfxsize=12.0cm \epsfbox{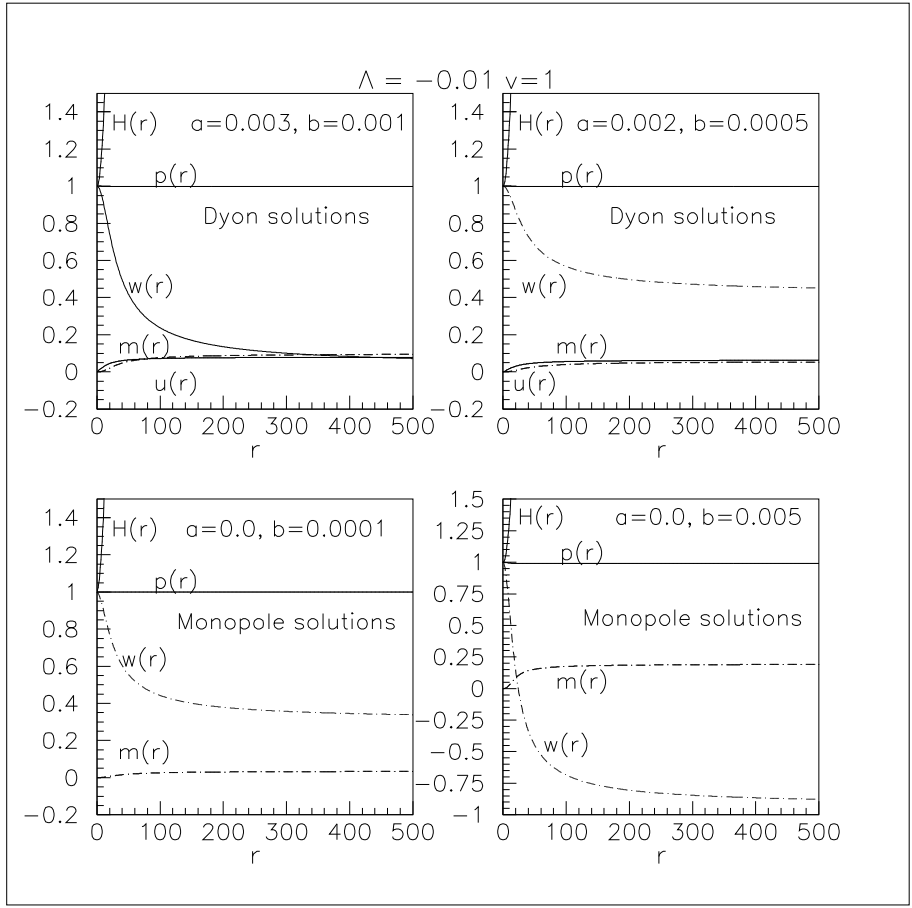}}
\vskip .5cm
\caption{Monopole and dyon solutions for  $\Lambda = -0.01$ and $v=1$.
$(a,b)=(0, 0.001)$ and $(0, 0.005)$ for the monopole solutions and
$(a,b)=(0.003, 0.001)$ and $(0.002, 0.0005)$ for the dyon solutions.}
\label{m_plot}
\end{figure}

As shown in fig.\ 1, the behavior of $m$ and $p$ is similar to that of
the asymptotically dS solutions  previously considered
\cite{VOLKOV}. In contrast, there exist solutions where $w$ has no
nodes, which are not seen in the asymptotically dS or Minkowski
cases.

\section{Dyon solutions}


Dyon solutions to the EYM equations for a given negative $\Lambda$
are found if we choose the adjustable shooting
parameter $a$ to be non-zero.
Fig. \ref{m_plot} shows  how the electric component, $u$, of the EYM
equations starts at zero and monotonically increases to some finite
value. The behavior of $w$, $m$ , $H$, and $p$ is similar to the
monopole solutions.

Again we find a continuum of solutions for a continuous set of
parameters $a$ and $b$, where $w$ crosses the axis an
arbitrary number of times depending these parameters. Also similar to
the monopole solutions is the existence of solutions where $w$ does not
cross the axis. This is in sharp contrast to the $\Lambda \ge 0$ case where
the dyon solutions are forbidden.  

\section{Stability of the monopole solutions}

The BK solutions and the dS-EYM
solutions are unstable \cite{ZHOU,Volkov4,BRODBECK}.
In contrast,  the AdS black hole solutions
\cite{WINSTANLEY} with $u=0$ and the monopole solutions without nodes\cite{BJORAKER}  are stable against
spherically symmetric linear fluctuations.

In order to derive the time dependent EYM equations, we use the most  general 
expression for the spherically symmetric SU(2) gauge fields 
in the singular gauge:
\begin{equation}
A=\frac{1}{2e} \, \Big\{ u\tau_3dt + \nu \tau_3 dr
+ (w\tau_1+\tilde{w}\tau_2)d\theta
+(\cot\theta\tau_3+w\tau_2-\tilde{w}\tau_1)\sin\theta d\phi\Big\} ,
\label{FIRST_A}
\end{equation}
where $\tau_i$ ($i=1$,$2$,$3$) are the usual Pauli matrices and $u$,
$\nu$, $w$ and $\tilde{w}$ depend on $r$ and $t$. 
The boundary conditions
$u=\nu=0$ and $w^2 + \tilde w^2 = 1$ at $r=0$ ensure regularity at the origin. 
Linearized equations when $u(t,r)=0$, for the gauge fields $\delta w(t,r)$, $\delta \tilde
w(t,r)$, $\delta \nu(t,r)$, $\delta p(t,r)$, and $\delta H(t,r)$
have been derived in the literature \cite{VOLKOV2}.   
Fluctuations decouple in terms of  $\delta w(t,r)$,   $\delta p(t,r)$, and $\delta H(t,r)$
which  form even-parity perturbations, and in terms of $\delta \tilde w(t,r)$ and 
$\delta \nu(t,r)$ which form odd-parity perturbations.  

The equation for parity-odd perturbations in  
$\beta = {r^2 p \delta\nu/ w}$, where $\beta (t,r) = e^{-i\omega
t}\beta (r)$, is  \cite{BJORAKER}
\beeq
\Bigg\{ - {d^2\over d\rho^2} + U_\beta(\rho) \Bigg\}  \beta
= \omega^2 \beta \next
U_\beta = \frac{H}{r^2p^2}(1+w^2)
+\frac{2}{w^2}\left(\frac{dw}{d\rho}\right)^2  
\label{U_nu}
\eneq
where ${d\rho/ dr} = {p/ H}$.
Volkov et al.\   showed that for the 
BK solutions there appear exactly $n$ negative eigenmodes
($\omega^2<0$) if $w$ has $n$ nodes \cite{Volkov4}.  Their argument
applies to the asymptotically AdS case without
modification.  One concludes that the
solutions with nodes in $w$ are unstable against parity-odd
perturbations.

For parity-even perturbations, where $\delta w(t,r) = e^{-i\omega t}
\delta w(r)$, we find the equation 
\beeq
\Bigg\{ - {d^2\over d\rho^2} + U_w(\rho) \Bigg\}  \delta w
= \omega^2 \delta w \next
U_w = \frac{H (3w^2-1)}{p^2r^2} 
+ 4 v \frac{d}{d\rho}\left(\frac{Hw'^2}{pr}\right) ~.
\label{U_w}
\eneq
Although the potential $U_w(\rho)$ is not positive definite, it is regular
in the entire range $0 \le \rho \le \rho_{\rm max}$.  The first term in 
$U_w$ becomes negative for $w^2 < 1/3$.  The 
Schr\"odinger equation (\ref{U_w}) was solved numerically \cite{BJORAKER}.
The potential for the solutions with
$(a,b)=(0,0.001)$, which has no node in $w$, has the lowest eigenvalue
$\omega^2$ of 0.028  and for $(a,b)=(0, 0.005)$,
which has one node, has the lowest eigenvalue of  0.023.  Therefore, these solutions are
stable against  parity-even perturbations, even if $w$ has one node, and  differ from the solutions
where $\Lambda \ge 0$ where the parity-even perturbations are
always unstable.

\section{Conclusion}

\begin{figure}[th]\centering
\leavevmode 
\mbox{
\epsfxsize=11.cm \epsfbox{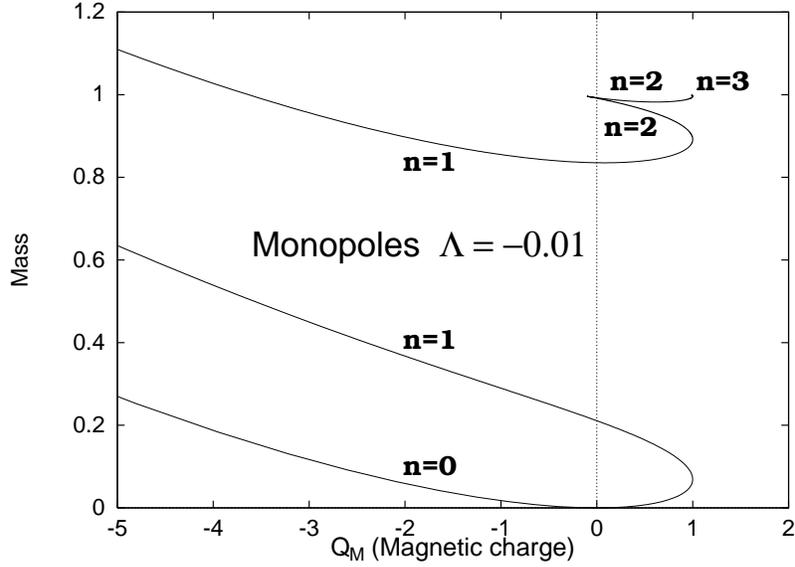}}
\vskip .5cm
\caption{The $Q_M$-$Mass$ plot of monopole solutions at $\Lambda=-0.01$.
The number of nodes, $n$, in $w(r)$ is also indicated.}
\end{figure}

A continuum of new monopole and dyon solutions to
the asymptotically AdS EYM equations have been found. 
In fig.\ 2 the spectrum of monopole solutions is plotted. The monopole
solutions in which the magnetic component $w(r)$ of the gauge fields
never vanishes, corresponding to the portion with $n=0$ in fig.\ 2,
were shown to be stable against linear perturbations.   
Solutions with $b \ge 0.7$  develop an event horizon, becoming black
holes. The end point of the $n=3$ portion of the curve in fig.\ 2 shows where the
solutions become black holes.

\leftline{\bf Acknowledgments}

This work was supported in part    by the U.S.\ Department of
Energy under contracts DE-FG02-94ER-40823 and DE-FG02-87ER40328.

\end{document}